\documentstyle[12pt,epsfig,rotating]{article}
\newcommand{\AmS}{{\protect\the\textfont2
  A\kern-.1667em\lower.5ex\hbox{M}\kern-.125emS}}
 \def\vf{\varphi} \def\la{\lambda}  
    \def\r#1{$^{[#1]}$}

  \def\vs{\vskip} \def\ni{\noindent}

\def\ifmath#1{\relax\ifmmode #1\else $#1$\fi}%

\def\rt{\ifmath{{\mathrm{t}}}}

\newcommand{\beqa}{\begin{eqnarray}} \newcommand{\eeqa}{\end{eqnarray}  }
\newcommand{\beqan}{\begin{eqnarray*}} \newcommand{\eeqan}{\end{eqnarray*}}
\newcommand{\beq}{\begin{equation}} \newcommand{\eeq}{\end{equation}  }
\def\cl{\centerline}

\begin{document}

\null{}\vskip -2.0cm
\hskip12cm{HZPP-9811}
\vskip-0.3cm

\hskip12cm{Dec. 12, 1998}
\vs1.5cm

\centerline{\Huge On the Factorial Moment Analysis of}
\vskip0.2cm
\centerline{\Huge High Energy Experimental Data}
\vskip0.2cm
\centerline{\Huge with Non-integer Partition Number}

\vskip1.5cm
\centerline{\Large Chen Gang}
\vskip0.5cm
\centerline{\small\it Department of Physics, Jingzhou Teacher's College,
Hubei 434104 China}
\vskip0.9cm
\centerline{\Large Liu Lianshou, 
\ \ \ \ \ \ \  Gao Yanmin}
\vskip0.5cm

\centerline{\small\it Institute of Particle Physics, Huazhong Normal University,
Wuhan 430079 China}

\vskip2cm
\begin{center}{\large ABSTRACT}\end{center}
\vskip0.5cm

\begin{center}\begin{minipage}{124mm}
{\small
   It is pointed out that in doing the factorial moment analysis with
non-integer partition $M$ of phase space, the influence of the 
phase-space variation 
of two- (or more-) particle correlations has to be considered 
carefully. In this paper this problem is studied and 
a systematic method is developed to minimize this influence.
The efficiency and self-consistency of this method are shown using
the data of 250 GeV/$c$ $\pi^+$p and K$^+$p collisions
from the NA22 experiment as example. }
\end{minipage}\end{center}

\newpage
\section{Introduction}
\vs-0.2truecm

    Since the pioneer work of Bialas and Peschanski\r{1}, the anomalous scaling of
factorial moments (FM) defined as 
\begin{equation}
F_q(\delta)=\frac{1}{M} \sum \limits_{m=1} ^M
\frac {\langle  n_m (n_m-1) \cdots (n_m-q+1) \rangle} {\langle n_m \rangle ^q},
\end{equation}   
has been searched extensively in high energy experiments\r{2}. 
In the above equation, $M$ is the partition number of a phase space
region $\Delta$ in consideration, $\delta=\Delta/M$ is the size of a sub-cell, 
$n_m$ is the number of (charged) particles falling into the $m$th sub-cell. 

After a long period of debate, it has been recognized that in studying the 
anomalous scaling of FM the anisotropy of phase space should be taken into
account\r{3}, i.e.  if dynamical fluctuation  does exist
in high energy hadron-hadron collisions then it should be   
anisotropic and the corresponding fractal is ``self-affine''\r{4}. 
This means that the anomalous scaling of FM can be observed when and only 
when the phase space is divided in an appropriate anisotropical way. 

Let the three phase space variables be denoted by $p_a, p_b, p_c$, the
corresponding shrinking ratios are $\la_a$, $\la_b$ and $\la_c$ 
respectively. The anisotropy (self-affinity) of dynamical fluctuation
can be charcterized by the so-called roughness or Hurst exponents\r{4} 
\begin{equation}
H_{ij}={\ln \la_i\over \ln \la_j},\ \ \ \ \ \ \  (i,j=a,b\quad \mbox{or} \quad
 a,c\quad
\mbox{or} \quad b,c),  \end{equation}   
\noindent with
\begin{equation}
\la_i \le \la_j, \quad \quad 0\le H_{ij}\le 1. \end{equation}   
These exponents can be deduced from the experimental data by fitting three 
one-dimensional second-order factorial-moments to the saturation curve\r{5}.
If self-affine fluctuations of multiplicity do exist in 
multiparticle production, exact scaling , i.e. a straight line 
in ln-FM versus ln$M$, should be observed when and only when the 
three-dimensional analysis is performed with the phase space divided 
anisotropically according to the value of these Hurst exponents.

In order to confront this assertion with experiment, an important point has
to be noticed, i.e. when $H\neq 1$ the partition numbers $M_i$ in the three
phase space directions, in general, cannot be integer numbers simultaneously.

It turns out that the case of 250 GeV/$c$ $\pi^+$p and K$^+$p collisions from 
NA22 experiments is special. The Hurst exponents determined from the 
one-dimensional second-order factorial-moments in these collisions are\r{6}
$H_{yp_\rt} = 0.48 \pm  0.06 \ ; \ 
H_{y\vf} = 0.47 \pm 0.06 \ ; \ 
H_{p_\rt\vf} = 0.99 \pm 0.01 \ .$
These Hurst exponents can be simply approximated by
$H_{yj}={1/2},\ (j=p_\rt, \vf),$ and $H_{p_\rt\vf}=1.$
In this approximation, the following integer values for $M_i$ can be used
simultaneously in higher-dimensional analysis: $M_y \ =1,2,3,\dots$ and 
$M_{p_\rt}=M_\vf=1,4,9,\dots$. 

In Ref. [6] this approximation has been used and the corresponding 3-D  
analysis has been preformed. The self-affine 3D ln$F_2$ vs. ln$M$ thus obtained
has been fitted to a straight line and the result is basically confirmative.
However, it can be seen from the above that the restriction of $M$ to integer 
values results in a 3-D plot with only 7 points.  Omitting the first point to 
get rid of the influence of transverse momentum conservation\r{7} only 6 points 
remain, which are unfortunately insufficient for a precise check of the
linearity of the plot.

In order to get more points in the self-affine 3-D ln-FM versus ln$M$ plot,
non-integer values of $M$ have to be used. In so doing an important problem
arises, i.e. the influence of the variation in phase space of two- 
(or more-) particle correlations has to be considered carefully. 
In this paper we will take this problem into account and try to 
develop a systematic method in minimizing the influence of the
deviation of two- (or more-) particle correlations from a constant behaviour. 
 
In section 2 the definition of factorial moments with non-integer partition
number $M$ is introduced and the problem of how to use it in real experimental
data analysis is discussed. It is shown that, due to the variation of 
two-particle correlations, a correction factor is needed.
How to extract this factor from experimental data is discussed in sections 3 
and 4.  Concluding remarks are given in section~5.

\section{Non-integer FM analysis}
\vs-0.2truecm

    For definiteness, let us consider the second order FM in one-dimensional 
phase space, e.g. rapidity $y$:
\begin{equation}
F_2(\delta y) = F_2(M)=\frac{1}{M} \sum \limits_{m=1} ^M
\frac {\langle  n_{m}  (n_{m}-1) \rangle} {\langle n_m \rangle ^2},
\end{equation}       
where $ M={\Delta y / \delta y}$ .

In the ideal case, factorial moments $F_q$ only depend on the bin width 
$\delta y$, but not on the position of the bin on the rapidity axis. 
If that is the case, the result of averaging over all the $M$ bins as
in Eq.(4) is equal to that of averaging over $N$ bins with $N\leq M$.
That is, $F_2(M) =F_2(N,M)$, where
\begin{equation}
 F_2(N,M)=\frac{1}{N} \sum \limits_{m=1} ^N
\frac {\langle  n_{m}  (n_{m}-1) \rangle} {\langle n_m \rangle ^2},
 \qquad (N=M-a, \ \ 0 \leq a<1). \end{equation}       
This equation can be used as the definition of an FM for any real value of
$M$\r{8}.

As is well known, even in the central region the rapidity distribution is
not flat. The shape of this distribution influences the scaling
behaviour of FM. At first, Fialkowski\r{9} has suggested to use a correction 
factor to minimize this influence. Later, the cumulant variable 
\begin{equation}
  x(y)= \frac{\int_{y_a}^y \rho(y) dy} {\int_{y_a}^{y_b} \rho(y) dy} 
\end{equation}      
was introduced\r{10}, which has a flat distribution and the Fialkowski
method of correction factor has been substituted.  It turns out, however, 
when the FM analysis with non-integer partition is concerned,
a similar factor has to be introduced again.

To see this, let $\Delta$ denote the phase space 
region in consideration, $\delta_m$ the $m$th bin, 
$\rho_1(y_1)$ and $\rho_2(y_1,y_2)$ the one- and two-particle
distribution functions, respectively. Then we have
$$ \rho_1(y_1)= {\langle n \rangle \over \langle n(n-1)\rangle}
\int_{\Delta}  \rho_2(y_1,y_2) dy_2; \eqno(7a)$$
$$ \langle n_m\rangle = \int_{\delta_m}\rho_1(y)dy
  = {\langle n \rangle \over \langle n(n-1)\rangle}
 \int_{\Delta} dy_2\int_{\delta_m} dy_1 \rho_2(y_1,y_2); \eqno(7b)$$
$$\langle n_m(n_m-1)\rangle= \int_{\delta_m}dy_2 \int_{\delta_m} 
   dy_1\rho_2(y_1,y_2). \eqno(7c)$$

After transforming to the cumulant variable, $\langle n_m\rangle$
becomes a constant, independent of $m$. Comparing the second and
third of the above equations, it can be seen
clearly that due to the difference in the integration region over $y_2$,
even though $\langle n_m\rangle$ is constant, $\langle n_m(n_m-1)\rangle$ 
is in general not constant. 

As an example, in Fig.1 are shown 
$f_2(m)=\langle n_m(n_m-1)\rangle/\langle n_m\rangle^2$ 
distributions in rapidity $y$ transformed into the cumulant variable.
Data are from $\pi^+$p and K$^+$p collisions at 250 GeV/$c$ (NA22).
The four figures correspond to $M=8,16,32,64$, respectively. It can be
seen that, although the cumulant variable has been used so that the
average distribution $\langle n_m\rangle$ is flat, the distribution of the
second-order factorial moment 
$f_2(m)=\langle n_m(n_m-1)\rangle/\langle n_m\rangle^2$  
is not flat.

Note that in the definition of $F_q$, equations (4) and (5), a horizontal
average has been taken. When the partition number $M$ is an integer, 
the horizontal average is over the whole region $\Delta$, cf. Eq.(4). 
The variation of 
$\langle n_m(n_m-1)\rangle/\langle n_m\rangle^2$  
is thus smeared out, and
no correction is needed. On the contrary, when $M$ is non-integer, the 
horizontal average is performed only over part of the region, cf. Eq.(5), 
and the influence of the variation of 
$\langle n_m(n_m-1)\rangle/\langle n_m\rangle^2$   
becomes essential. This influence is exhibited clearly
in Fig.2\footnote{As in the case of integer $M$, the errors of the data 
points are correlated. This problem has been discussed in Ref.[6].}. In 
these figures, ln$F_2$ versus ln$M$ for $y$, $p_{\rm t}$ and $\vf$
are shown with both integer (full circles) and non-integer (open circles)
values of $M$. All these plots show a ``sawtooth'' shape. The ``sawteeth'' come
from non-integer $M$. They lie above the smooth curve of integer $M$ in the
cases of $y$ and $p_{\rm t}$, but reach from above the smooth curve of integer 
$M$ to below it in the case of $\vf$.

In the following we will introduce a correction factor to minimize 
the influence of the variation of 
$\langle n_m(n_m-1)\rangle/\langle n_m\rangle^2$   
and remove 
the ``sawteeth'' from the ln$F_2$ versus ln$M$ plots for real $M$.

\section{Correction factor for the
$\langle n_m(n_m-1)\rangle/\langle n_m\rangle^2$   
distribution}
\vs-0.2truecm

From the  definition of $F_2(N,M)$, Eq.(5), it can be seen that
when $M$ is non-integer ($M=N+a, 0\leq a<1)$, only $N$ bins are included in 
the horizontal average, i.e. only $r=N/M$ of the whole 
region $\Delta$ has been taken into account. The result is therefore
influenced by the variation of 
$\langle n_m(n_m-1)\rangle/\langle n_m\rangle^2$.
We try to minimize this influence by introducing a correction factor 
$R(r)$ and define
$$ F_2(M)=\frac{1} {R(r)} \left(\frac{1}{N} \sum \limits_{m=1} ^N
\frac {\langle  n_{m}  (n_{m}-1) \rangle} {\langle n_m \rangle ^2}\right).
\eqno(8) $$

In order to extract the correction factor $R(r)=R(N/M)$ from the experimental
data, let us choose two integers $N'$ and $M'$, satisfying
  $$ \frac{N'}{M'} \approx \frac{N}{M} = r. \eqno(9)  $$
One is then ready to calculate $F_2(N',M')$  according to Eq.(5).

Let us call the ratio 
 $$ C(N',M')= \frac {\frac{1}{N'} \sum \limits_{m=1} ^{N'}
    \langle  n_{m}  (n_{m}-1) \rangle/\langle n_m\rangle^2}   
 {\frac{1}{M'} \sum \limits_{m=1} ^{M'}
    \langle  n_{m}  (n_{m}-1) \rangle/\langle n_m\rangle^2} \eqno(10)  $$
of $F_2(N',M')$ to $F_2(M')$ correction matrix. It is the ratio of the FM 
averaging only over $N'/M'$ of the whole region to the FM averaging 
over the whole region. Having Eq.(9) in mind, it is reasonable, therefore, 
to take $C(N'.M')$ as the approximate value of the correction factor 
$R(r)=R(N/M)$.

In the left column of Fig.3 are shown the correction matrix $C(N,M)$ 
with $N$ and $M$ both taking integer values as function of $N/M$ for 
$M=3,4,\dots,40$. It can be seen from the figures that all the points 
for different $M$'s lie in a narrow band, i.e. $C(N,M)$ is mainly a function
of $r=N/M$ and is only weakly dependent on $M$. This indicates that the 
correction factors $R_y(r)$, $R_{p_{\rm t}}(r)$ and $R_\vf(r)$ 
for non-integer $M$ can be obtained from the interpolation of the
corresponding $C(N,M)$'s. 

It turns out that in order to eliminate
the ``sawteeth'' in the ln$F_2$ versus ln$M$ plot appropriately, the 
interpolation curve should be chosen in  the following way\r{11}: 
When the ``sawteeth'' in the ln$F_2$ versus ln$M$ plot lie
above the smooth curve of integer $M$, as in the case of $y$ and $p_{\rm t}$, 
the upper boundary of the $C(N,M)$ band has to be used for the interpolation; 
while when the ``sawteeth'' in the
ln$F_2$ versus ln$M$ plot reach from above the 
smooth curve of integer $M$ to below it, as in the case of $\vf$, the middle 
of the $C(N,M)$ band has to be used, cf. the dotted lines in the right 
column of Fig.3. 

Using these interpolation curves to get the correction factors, the
1-D FM's can be corrected according to Eq.(8). The results are shown in
Fig.4. The sawteeth seen in Fig.2 are largely eliminated.

\section{Higher-dimensional Correction}
\vs-0.2truecm

Having obtained the correction factors for one-dimensional FM's with
non-integer $M$, let us turn to the correction of higher-dimensional FM's.

First, consider the 2-D case.
Let $M_1$, $M_2$ be the non-integer partitions in directions 1 and 2,
respectively, $N_1$, $N_2$ their integer parts. ($M_1=N_1+a_1$, $M_2=N_2+a_2$,
$0\leq a_1<1$, $0\leq a_2<1)$. The definition of a non-integer FM, Eq.(8),
is extended to 
  $$ F_2(N_1,M_1;N_2,M_2)=\frac{1}{N_1N_2} \sum \limits_{m_1=1} ^{N_1}
                       \sum \limits_{m_2=1} ^{N_2}
            \frac {\langle  n_{m_1,m_2}  (n_{m_1,m_2}-1) \rangle} 
                        {\langle n_{m_1,m_2} \rangle ^2}.  \eqno(11) $$
The average is taken over the area $N_1\cdot N_2$, shown 
as blank area in Fig.5$a$. Since the distribution of 
${\langle  n_{m_1,m_2}  (n_{m_1,m_2}-1) \rangle}$ is 
in general not flat, a correction 
should be applied to take the shaded area in Fig.5$a$ into account.

The one-dimensional correction factors $R_1$, $R_2$ of directions 1 and 2
can take care of the areas $\Delta S_1=N_2(M_1-N_1)$ and 
$\Delta S_2=N_1(M_2-N_2)$, respectively. However, the crossing area 
$\Delta S_{12}=(M_1-N_1)(M_2-N_2)$ has still to be considered. We evaluate it
using an area counting method. Let the area ratios be
  $$  b_1={\Delta S_{12} \over \Delta S_2} ={M_1-N_1\over N_1} \ \ ; \ \  
      b_2={\Delta S_{12} \over \Delta S_1} ={M_2-N_2\over N_2}.  \eqno(12) $$
The correction factor due to $\Delta S_{12}$ can be written as 
 $$ R_{12} =R_1^{b_2/2}\cdot R_2^{b_1/2}. $$ 
Noting that $R_1\approx 1$, $R_2\approx 1$, we have approximately
  $$  R_{12} = 1 + {1 \over 2} [b_1(R_2-1)+ b_2(R_1-1)].  \eqno(13) $$
The correction factor $R^{(2)}$ for 2-dimensional FM is then
  $$ R^{(2)} =R_1 \cdot R_2 \cdot R_{12} , \eqno(14) $$
  $$ F_2(M_1,M_2) = {1 \over R^{(2)} } F_2(N_1,M_1;N_2,M_2). \eqno(15) $$

In the left column of Fig.6 are shown the 2-D FM's with real partition number $M$
without correction. A strong ``sawtooth'' effect can be seen. Using 
correction factors obtained as above, Eq.(13-15), the large sawteeth
disappear, cf. the right column of Fig.6. 

Similar considerations lead to  
the correction factor for the 3-dimensional case, cf. Fig.5$b$:
  $$ F_2(M_1,M_2,M_3) = {1 \over R^{(3)} } 
       F_2(N_1,M_1;N_2,M_2;N_3,M_3). \eqno(16) $$
 $$ R^{(3)} = R_1\cdot R_2 \cdot R_3 \cdot R_{12} \cdot R_{23} \cdot
                 R_{31} \cdot R_{123}, \eqno(17) $$
 $$ R_{12} = 1 + {1\over 2} [b_1(R_2-1) + b_2(R_1-1) ]; $$
 $$ R_{23} = 1 + {1\over 2} [b_2(R_3-1) + b_3(R_2-1) ]; $$
 $$ R_{31} = 1 + {1\over 2} [b_3(R_1-1) + b_1(R_3-1) ]; $$
 $$ R_{123} = 1 + {1\over 3} [ b_1b_2(R_3-1) + b_2b_3(R_1-1) 
                + b_3b_1(R_2-1)]. $$
 $$b_1={\Delta V_{12} \over \Delta V_2}=
              {\Delta V_{31} \over \Delta V_3}={M_1-N_1 \over N_1}; $$
 $$b_2={\Delta V_{12} \over \Delta V_1}=
              {\Delta V_{23} \over \Delta V_3}={M_2-N_2 \over N_2}; $$
 $$b_3={\Delta V_{31} \over \Delta V_1}=
              {\Delta V_{23} \over \Delta V_2}={M_3-N_3 \over N_3}. $$
The resulting 3D plots before and after correction are shown in Fig.7.

\section{Discussion}
\vs-0.2truecm

In this paper we have developed a systematic method for an FM analysis
with real (integer and non-integer) partition $M$.
Correction factors are introduced 
in order to minimize the influence of the variation of 
$f_2(m)=\langle n_m(n_m-1)\rangle/\langle n_m\rangle^2$. 
The corrected results for non-integer $M$ lie on
smooth curves together with the integer $M$ points.

Note that only one-dimensional correction factors need to be extracted
from the experimental data. In higher-dimensional correction
the correction factors determined in one dimension have been used 
together with a simple geometrical consideration. This confirms the 
self-consistency of the method.

\vskip1cm
\ni 
{\Large \bf Acknoledgement}
\vskip0.5cm

The authors are grateful to Prof. W. Kittel and the NA22 Collaboration for
allowing them to use NA22 data as example of the proposed
method. Valuable comments on the draft from W. Kittel are highly appreciated.
This work is supported in part by NSFC and the Science Fund of Hubei 
Province. We further thank the National Commission of Science and Technology
of China (NCSTC) and the Royal Academy of Science of the Netherlands (KNAW)
for support within the program Joint Research of China and the Netherlands 
under project number 97CDP004.

\newpage

\noindent{\bf\large\bf Figure captions}
\vskip0.1cm

\noindent
Fig.1 \ $f_2(m)=\langle n_m(n_m-1)\rangle/\langle n_m\rangle^2$,  
the distribution in 
rapidity $y$ transformed into the cumulant variable.

\noindent Fig.2 \ The one-dimensional plots of ln$F_2$ versus ln$M$
for real $M$ (from $M =1$ to $M=3.6$) as defined in Eq.(5). The full circles 
correspond to integer $M$, while the open circles correspond to non-integer 
$M$. 

\noindent Fig.3 \ The correction matrix $C(N,M)$ as a function of $N/M$ for
$M=3$ -- $40$.

\noindent Fig.4 \ The one-dimensional plots of ln$F_2$ versus ln$M$
for real $M$ (from $M =1$ to $M=3.6$) as defined in Eq.(5) corrected
according to Eq.(8). The full circles 
correspond to integer $M$, while the open circles correspond to non-integer 
$M$. 

\noindent Fig.5 \ Schematic plot of the phase-space area (volumn) used in the 
average of a two- (three-) dimensional FM for non-integer partition $M$. 
The shaded areas (volumns) are those which have to be accounted for in the 
correction factors.

\noindent Fig.6 \ The two-dimensional plots of ln$F_2$ versus ln$M$. The 
left column are the results of real $M$ as defined in Eq.(5). The right 
column are the results after correction.

\noindent Fig.7 \ The three-dimensional plots of ln$F_2$ versus ln$M$. The 
left column is the result of real $M$ as defined in Eq.(5). The right 
column is the result after correction.

\newpage

\begin{picture}(260,240)
\put(-90,-440)
{\epsfig{file=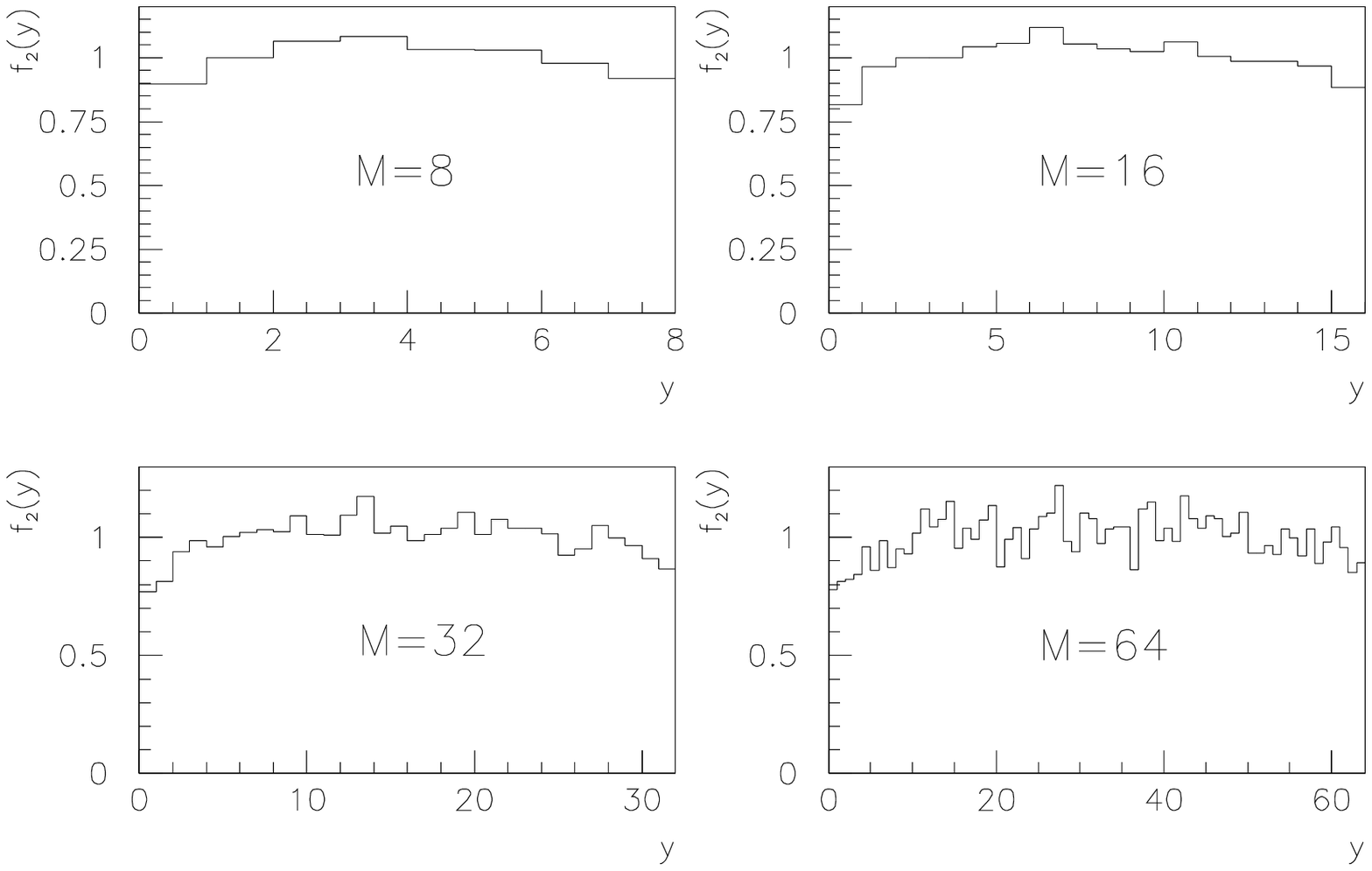,bbllx=0cm,bblly=0cm,
	   bburx=8cm,bbury=6cm}}


\end{picture}
\vskip5.8cm
 \cl{\Large Fig.1  }

\newpage

\begin{picture}(260,240)
\put(-90,-440)
{\epsfig{file=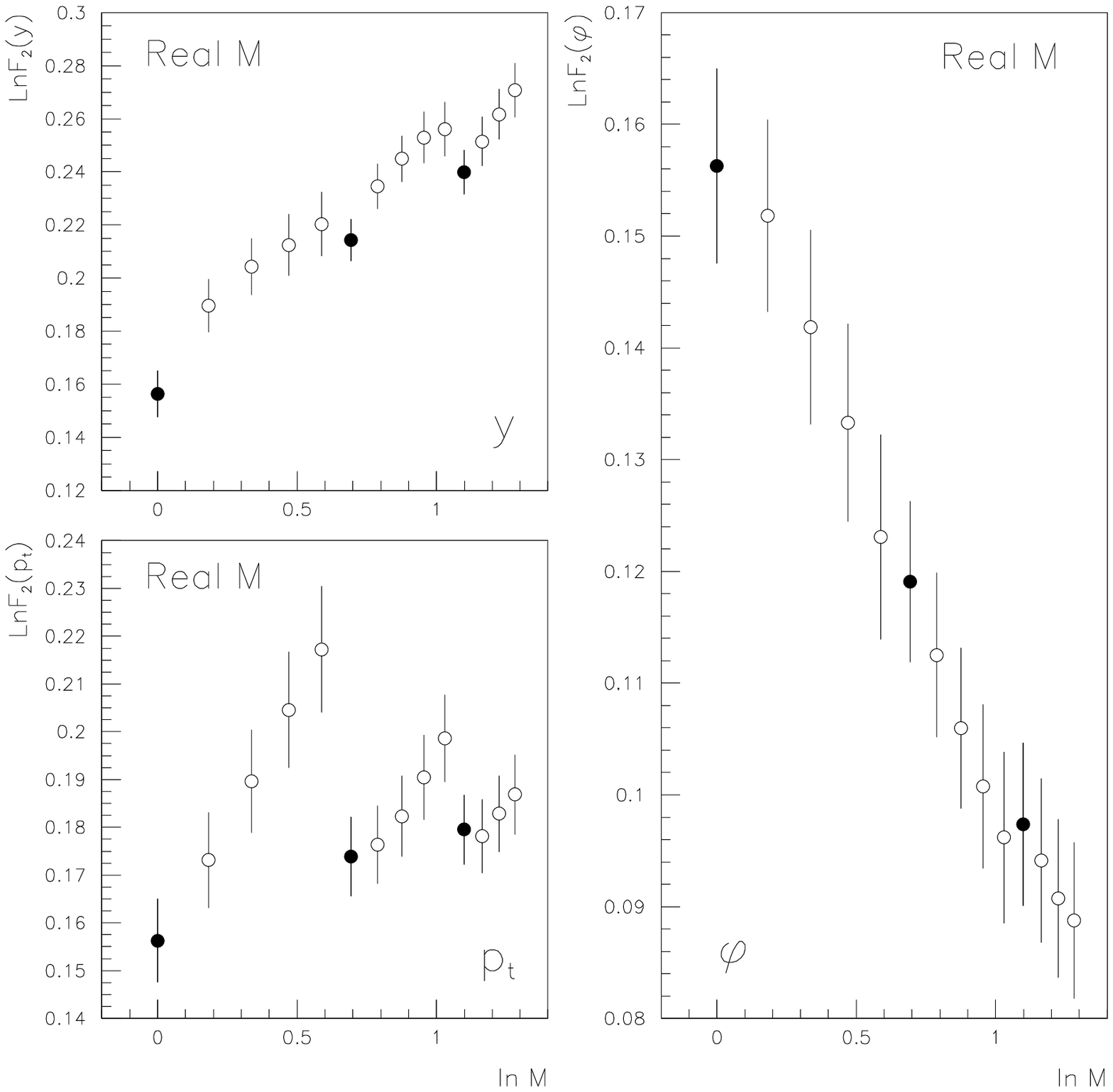,bbllx=0cm,bblly=0cm,
	   bburx=8cm,bbury=6cm}}


\end{picture}
\vskip10.5cm
 \cl{\Large Fig.2  }

\newpage

\begin{picture}(260,240)
\put(-90,-440)
{\epsfig{file=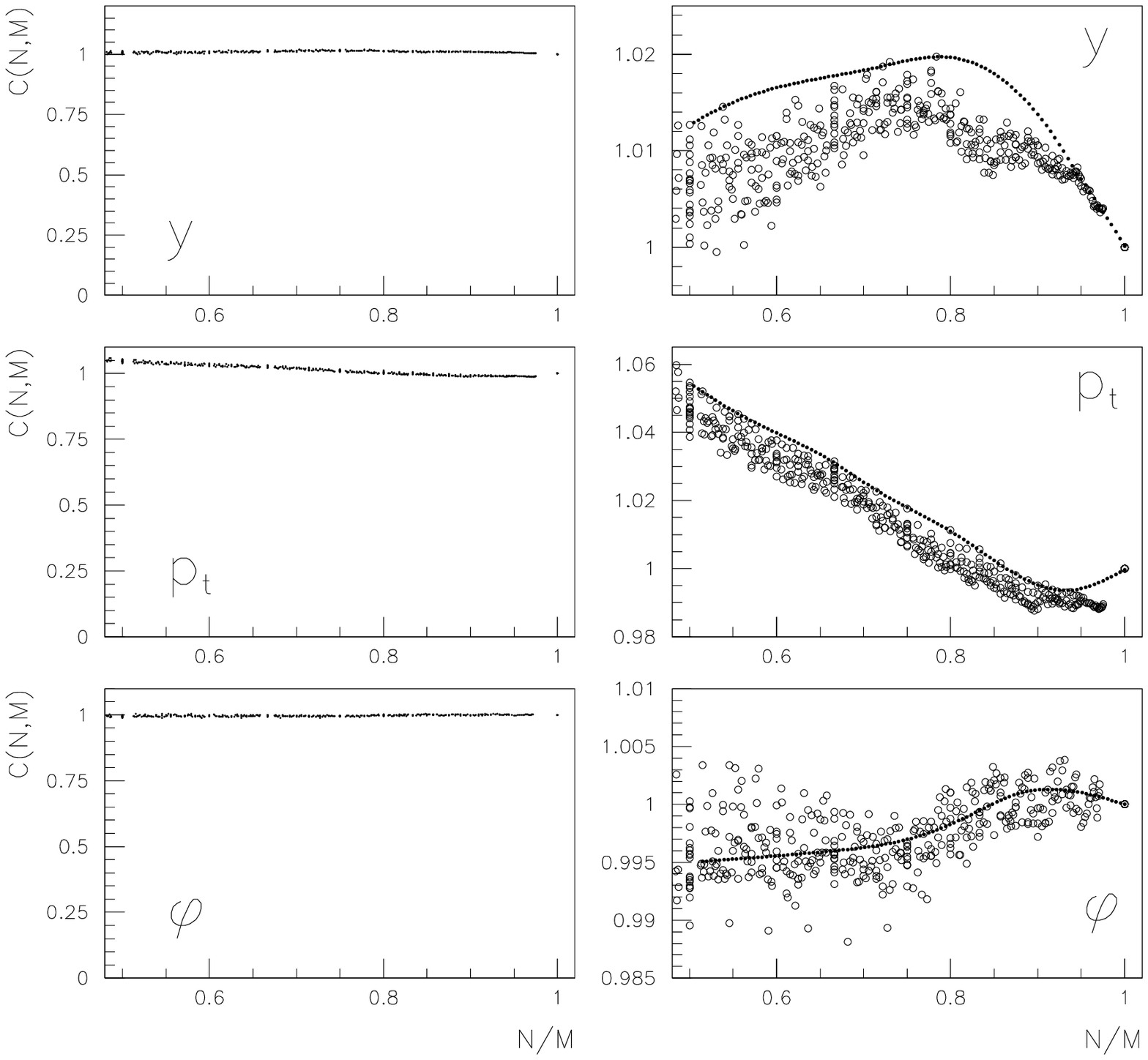,bbllx=0cm,bblly=0cm,
	   bburx=8cm,bbury=6cm}}


\end{picture}
\vskip10.5cm
 \cl{\Large Fig.3  }

\newpage

\begin{picture}(260,240)
\put(-90,-440)
{\epsfig{file=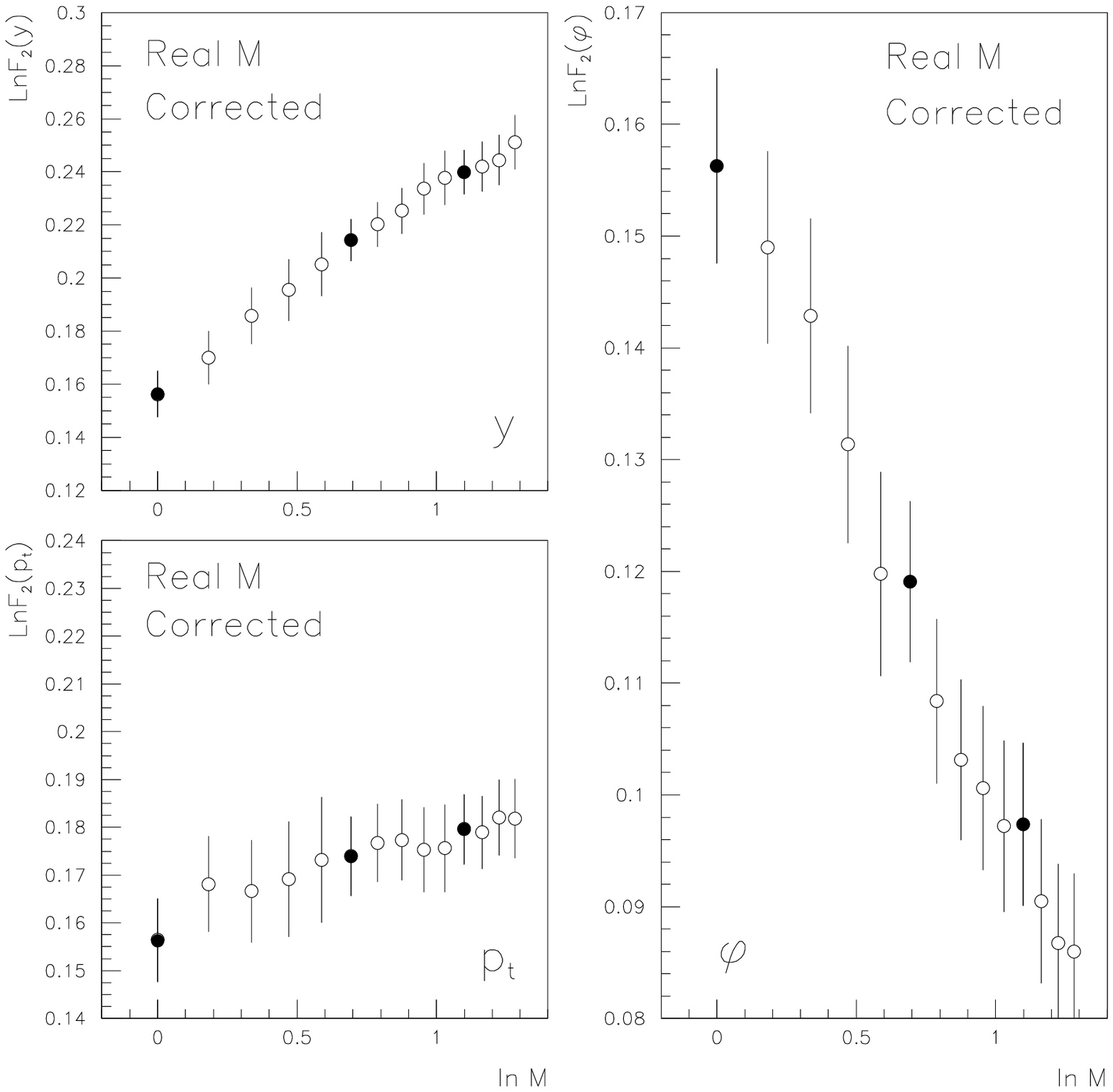,bbllx=0cm,bblly=0cm,
	   bburx=8cm,bbury=6cm}}


\end{picture}
\vskip10.8cm
 \cl{\Large Fig.4  }

\newpage

\begin{picture}(260,240)
\put(-90,-640)
{\epsfig{file=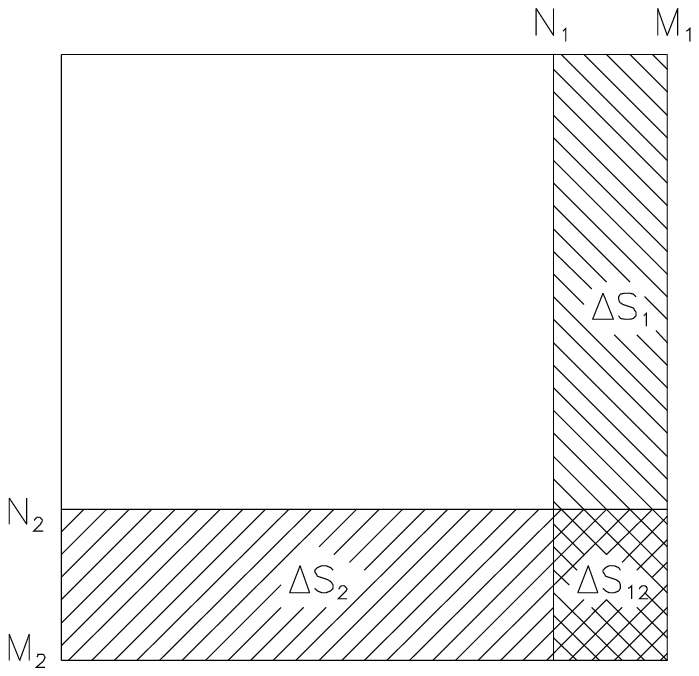,bbllx=0cm,bblly=0cm,
	   bburx=8cm,bbury=6cm}}


\end{picture}
\begin{picture}(260,240)
\put(-90,-640)
{\epsfig{file=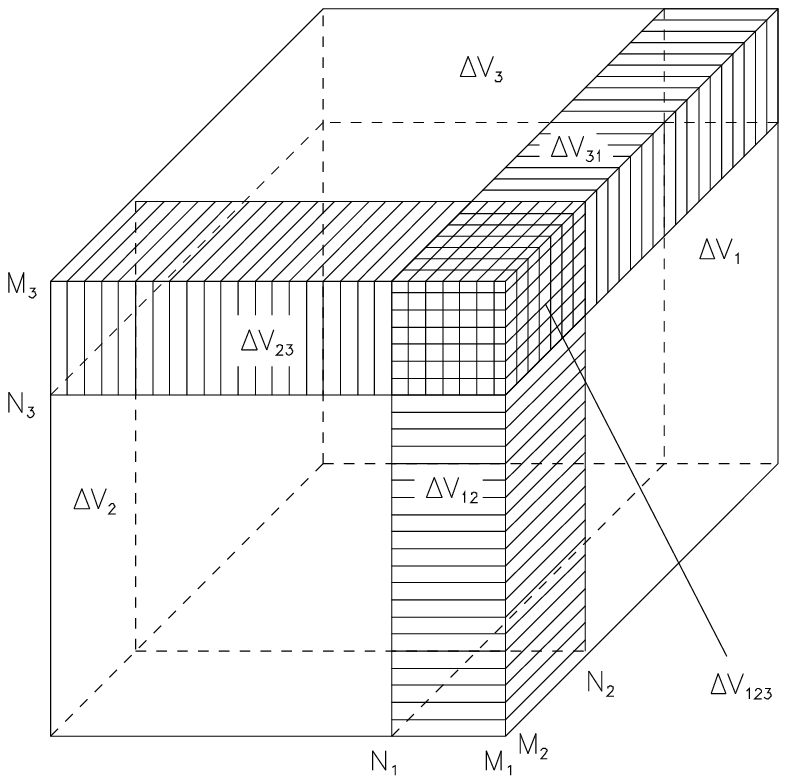,bbllx=0cm,bblly=0cm,
	   bburx=8cm,bbury=6cm}}


\end{picture}
\vskip4.5cm
 \cl{\huge $(a)$ \hskip7cm $(b)$ }
\vskip1.5cm
 \cl{\Large Fig.5  }

\newpage

\begin{picture}(260,240)
\put(-160,-440)
{\epsfig{file=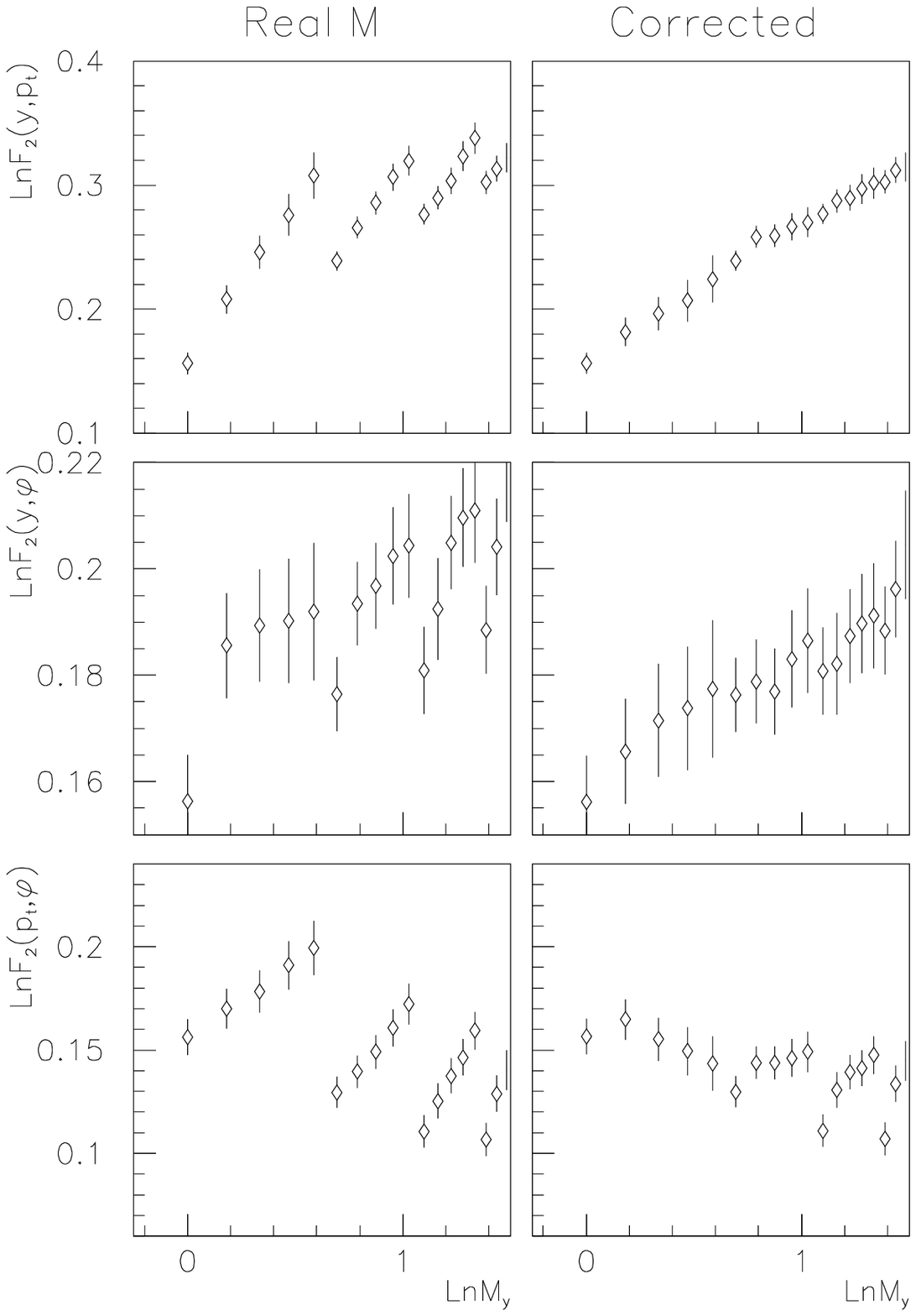,bbllx=0cm,bblly=0cm,
	   bburx=8cm,bbury=6cm}}


\end{picture}
\vskip10.8cm
 \cl{\Large Fig.6  }
\newpage

\begin{picture}(260,240)
\put(-90,-440)
{\epsfig{file=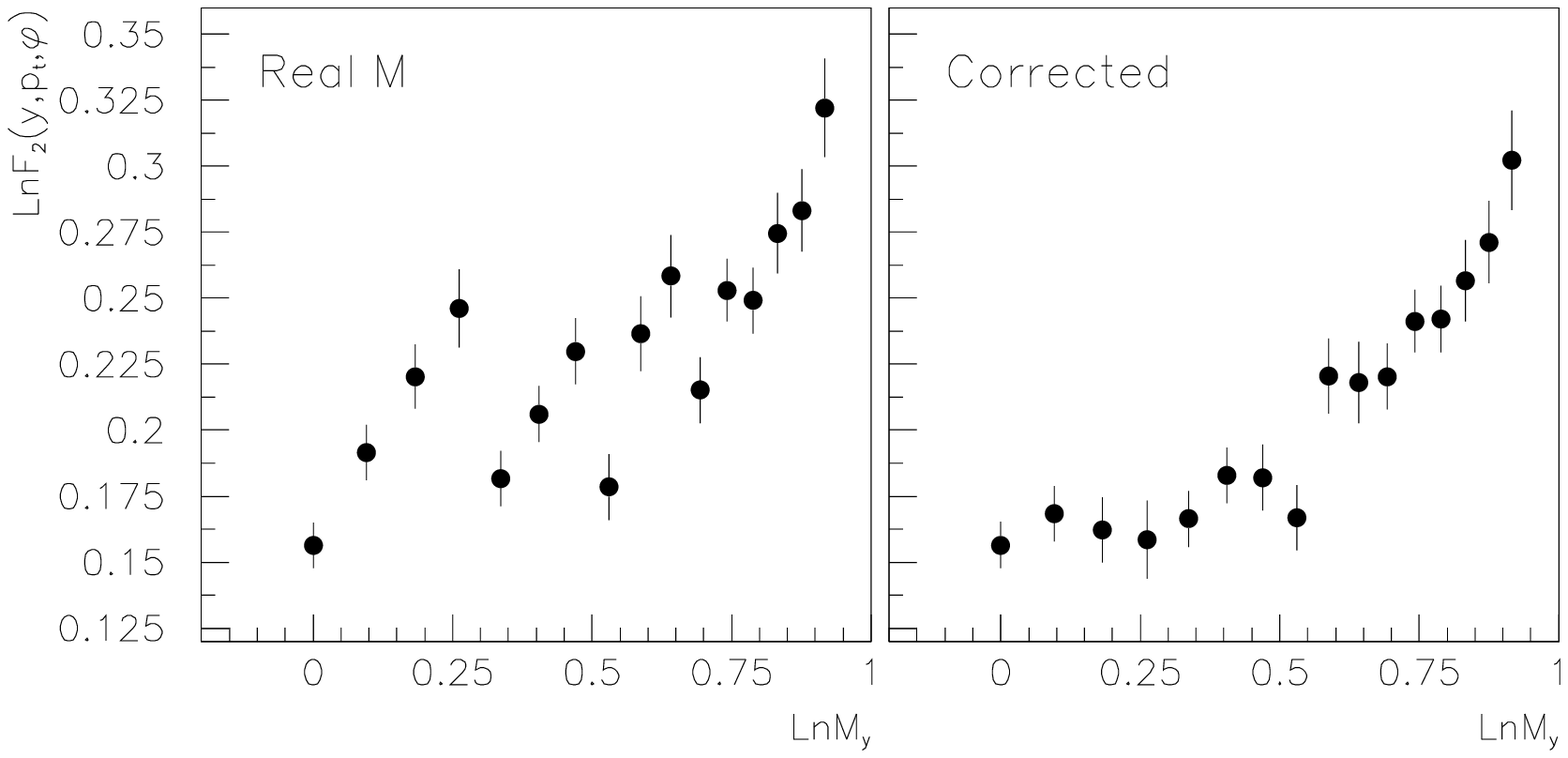,bbllx=0cm,bblly=0cm,
	   bburx=8cm,bbury=6cm}}


\end{picture}
\vskip3.0cm
 \cl{\Large Fig.7  }

\end{document}